\documentclass[a4paper]{report}
\usepackage[utf8]{inputenc}
\usepackage[T1]{fontenc}
\usepackage{RJournal}
\usepackage{amssymb, amsmath, amsthm, mathtools}
\usepackage{booktabs}

\usepackage{bm, color}
\usepackage{threeparttable}

\newcommand{\class}[1]{`\code{#1}'}
\newcommand{\fct}[1]{\code{#1()}}

\newcommand{\e}{{\rm e}}

\newcommand{\dd}{{\rm d}}

\newcommand{\ind}{\mathbb{I}}

\newtheorem{definicao}{Definition} 

\begin{document}

\sectionhead{Contributed research article}
\volume{XX}
\volnumber{YY}
\year{20ZZ}
\month{AAAA}

\begin{article}
\title{Comprehensive Regression and Diagnostics for Non-Negative Data Using the BCSreg Package}
\author{by Francisco F. Queiroz, Rodrigo M. R. de Medeiros}

\maketitle

\abstract{
Continuous positive data characterized by high skewness and heavy tails frequently arise in applied statistics. In other applications, these characteristics are accompanied by a point mass at zero, resulting in a non-negative response with a mixed discrete-continuous distribution. Standard regression models often fail to capture these complex features adequately, requiring more flexible approaches. In this paper, we introduce the \CRANpkg{BCSreg} package for R, which provides a comprehensive and unified computational framework for fitting Box-Cox symmetric and log-symmetric regression models for positive continuous data and their zero-adjusted extensions for mixed non-negative data. These broad classes of models accommodate varying degrees of skewness and tail-heaviness while allowing the parameters to be interpreted directly on the original scale of the data. Through a user-friendly multi-part formula interface, the \CRANpkg{BCSreg} package allows practitioners to simultaneously specify regression structures for the scale parameter (which is proportional to the quantiles of the response), the relative dispersion, and, when appropriate, the probability of zero occurrences. Furthermore, the package provides a complete suite of diagnostic tools specifically tailored to these classes of models, including randomized quantile residuals, simulated envelopes, and influence diagnostics. The package's features and capabilities are illustrated through applications to real data.
}

\section{Introduction}

Positive continuous data frequently appear in studies across various fields, including economics, climatology, and public health. Typical examples include household expenditures, wind speed, and medical costs, where the primary objective is to model the response variable as a function of a set of explanatory variables. In other applications, however, a substantial proportion of the response observations are exactly zero, giving rise to a mixed discrete-continuous distribution on the non-negative real line. Standard regression models for positive continuous data include gamma, inverse Gaussian, log-normal, and log-symmetric regression models \citep{vanegas2015}. When exact zeros are present, zero-adjusted versions of these models, as well as Tweedie regression models \citep{bonat2017}, are commonly employed. In R, many of these approaches are implemented in packages such as \CRANpkg{gamlss} \citep{stasinopoulos2008}, \CRANpkg{gamlss.inf} \citep{enea2025}, \CRANpkg{glmmTMB} \citep{brooks2017}, and \CRANpkg{ssym} \citep{vanegas2016ssym}. Nevertheless, these probability models often lack the flexibility required to adequately capture complex distributional features, such as pronounced skewness and heavy tails.

The class of Box-Cox symmetric (BCS) distributions \citep{ferrari2017} provides a flexible framework for modeling positive data with varying degrees of skewness and tail heaviness. Built upon the Box-Cox normal distribution \citep{cole1992}, the BCS framework assumes that a generalized Box-Cox transformation of the positive response variable follows a truncated symmetric distribution. Recently, \citet{demedeirosbcs} formalized the general class of BCS regression models and introduced its zero-adjusted extension (ZABCS) to properly accommodate data exhibiting a probability mass at zero. 

Although some specific members of the BCS class, such as the Box-Cox normal, Box-Cox t, and Box-Cox power exponential regressions, can be fitted using the \CRANpkg{gamlss} package \citep{rigby2004, rigby2006, stasinopoulos2008}, with their zero-adjusted extensions available through the \CRANpkg{gamlss.inf} package, a unified implementation encompassing the broader BCS and ZABCS regression frameworks is still lacking. The \CRANpkg{BCSreg} package fills this gap by providing a comprehensive suite of tools for fitting, evaluating, and diagnosing BCS and ZABCS regression models. A key feature of the package is the unified implementation of four regression model classes: the BCS \citep{demedeirosbcs} and log-symmetric \citep{vanegas2015} classes for positive continuous data, and their zero-adjusted counterparts, namely the ZABCS \citep{demedeirosbcs} and zero-adjusted log-symmetric \citep{cunha2024} classes, for mixed non-negative data. Using a multi-part formula interface, \CRANpkg{BCSreg} allows users to simultaneously specify regression structures for the scale parameter, the relative dispersion, and the probability of zero occurrences. In addition to maximum likelihood estimation, the package implements diagnostic methods specifically tailored to these models, including randomized quantile residuals, simulated envelopes, and influence diagnostics.

The remainder of this paper is organized as follows. Section \ref{sec:BCSmodel} reviews the theoretical definitions of the BCS and ZABCS distributions, as well as their associated regression structures. Section \ref{sec:implementation} describes the computational implementation and the main functions available in the \CRANpkg{BCSreg} package. Section \ref{sec:illustrations} illustrates the practical use of the package using a real datasets, demonstrating model specification, hypothesis testing for the skewness parameter, and residual analysis. Finally, Section \ref{sec:summary} presents the concluding remarks.

\section{Box-Cox symmetric regression models and its zero-adjusted extension}\label{sec:BCSmodel}

In what follows, we present the BCS and ZABCS distributions and their respective regression models.

\subsection{Box-Cox symmetric distributions}

The Box-Cox symmetric (BCS) distributions \citep{ferrari2017} are defined from a transformation of a positive continuous random variable whose distribution is truncated standard symmetric with a probability density function proportional to $r(z^2)$, where $r:[0,\infty) \rightarrow [0, \infty)$ satisfies $\int_0^\infty u^{-1/2}r(u)\mathrm{d}u = 1$. The function $r(\cdot)$ is called the density generating function (DGF). We say that a positive continuous random variable $Y$ follows a Box-Cox symmetric distribution with parameters $\mu > 0$, $\sigma > 0$, and $\lambda \in \R$ if
\begin{equation*}
Z \coloneqq T(Y; \mu, \sigma, \lambda) = \left\{
\begin{array}{ll}
\dfrac{1}{\sigma \lambda} \left\{\left(\frac{Y}{\mu}\right)^\lambda - 1 \right\}, & \mbox{ if } \lambda \neq 0, \\
\dfrac{1}{\sigma} \log\left(\frac{Y}{\mu}\right), & \mbox{ if } \lambda = 0,
\end{array}
\right.
\end{equation*}
has a standard symmetric distribution truncated at $\R\backslash A(\sigma, \lambda)$ (i.e., the support of the truncated distribution is $A(\sigma, \lambda)$), where
\begin{equation*}
A(\sigma, \lambda) = \left\{
\begin{array}{cl}
\left(-\frac{1}{\sigma \lambda}, \infty\right), & \mbox{ if } \lambda > 0, \\
\left(-\infty, -\frac{1}{\sigma \lambda}\right), & \mbox{ if } \lambda < 0, \\
(-\infty, \infty), & \mbox{ if } \lambda = 0.
\end{array}
\right.
\end{equation*}
We write $Y \sim \text{BCS}(\mu, \sigma, \lambda; r)$. The DGF $r(\cdot)$ may depend on an extra parameter, denoted here by $\zeta$. The distribution of $Y$ depends on the chosen DGF. For instance, if $Z$ has a truncated standard normal distribution, then $Y$ has a Box-Cox normal (BCNO) distribution; if $Z$ has a truncated standard Student-$t$ distribution with $\zeta$ degrees of freedom, then $Y$ has a Box-Cox $t$ (BCT) distribution with extra parameter $\zeta$.

The probability density function (PDF) of $Y \sim \text{BCS}(\mu, \sigma, \lambda; r)$ is
\begin{equation*}
f(y; \mu, \sigma, \lambda) = \begin{cases} 
\frac{y^{\lambda-1}}{\mu^\lambda \sigma} \frac{r(z^2)}{R\left(\frac{1}{\sigma |\lambda|}\right)}, & \text{if } \lambda \neq 0, \\ 
\frac{1}{y\sigma} r(z^2), & \text{if } \lambda = 0, 
\end{cases}
\end{equation*}
for $y > 0$, where $z = T(y; \mu, \sigma, \lambda)$ and $R(x) = \int_{-\infty}^x r(u^2)\mathrm{d}u$. The cumulative distribution function (CDF) of $Y$ is $F(y; \mu, \sigma, \lambda) = R(z) / R(-1/(\sigma \lambda))$ if $\lambda < 0$, $R(z)$ if $\lambda = 0$, and $[R(z) - R(-1/(\sigma \lambda))] / R(1/(\sigma \lambda))$ if $\lambda > 0$.

To accommodate exact zeros, the zero-adjusted Box-Cox symmetric (ZABCS) distributions are constructed from a mixture of a degenerate random variable at zero and a BCS continuous random variable \citep{demedeirosbcs}. A random variable $\mathcal{Y} \sim \text{ZABCS}(\alpha, \mu, \sigma, \lambda; r)$ has the CDF
\begin{equation*}
F^{(0)}(y; \alpha, \mu, \sigma, \lambda) =  \alpha \mathbb{I}(y \geq 0) + (1-\alpha) F(y; \mu, \sigma, \lambda),
\end{equation*}
where $\alpha \in (0,1)$ is the probability of observing a zero, $\mathbb{I}(\cdot)$ is the indicator function, and $F(y; \mu, \sigma, \lambda)$ is the CDF of the continuous BCS part. Particular cases of this extension include the zero-adjusted BCNO (ZABCNO) and zero-adjusted BC$t$ (ZABCT) distributions. 

The BCS and ZABCS distributions possess several interesting properties. For instance, the parameters $\mu$, $\sigma$, and $\lambda$ represent, respectively, a scale parameter (corresponding to the median when $\lambda=0$), the relative dispersion, and the skewness of the BCS distributions, as well as of the conditional distribution of $\mathcal{Y}$ given $\mathcal{Y}>0$ for the ZABCS models. Notably, the ZABCS distributions are not absolutely continuous due to a point mass at zero, where $\alpha = \mathbb{P}(\mathcal{Y}=0)$ represents the probability of observing a zero. When $\lambda = 0$, the BCS and ZABCS models reduce to the log-symmetric \citep{vanegas2016} and zero-adjusted log-symmetric distributions \citep{cunha2024}, respectively. Furthermore, these models offer remarkable flexibility in representing right-tail behaviors—ranging from light to extremely heavy tails—which are controlled by $\lambda$ and the extra parameter $\zeta$. For further details, see \cite{demedeirosbcs}.

\subsection{Box-Cox symmetric and zero-adjusted Box-Cox symmetric regression models}\label{sec:BCS_ZABCS_regression}

The BCS and ZABCS regression models are defined as follows. 

\begin{definicao}[Box-Cox symmetric regression models].
Let $Y_1, \ldots, Y_n$ be independent random variables, where $Y_i \sim \text{BCS}(\mu_i, \sigma_i, \lambda; r)$ for $i=1,\ldots,n$. The BCS regression models are defined by the systematic components
\begin{equation*}
d_1(\mu_i) = \bm{x}_i^{\top}\bm{\beta} \quad \text{and} \quad d_2(\sigma_i) = \bm{s}_i^{\top}\bm{\tau},
\end{equation*}
where $\bm{\beta} \in \mathbb{R}^p$ and $\bm{\tau} \in \mathbb{R}^q$ are unknown parameter vectors ($p + q + 1 < n$), and $\bm{x}_i$ and $\bm{s}_i$ are known covariate vectors associated with full-rank model matrices $\textbf{X}=[\bm{x}_1,\ldots,\bm{x}_n]^\top$ and $\textbf{S}=[\bm{s}_1,\ldots,\bm{s}_n]^\top$, respectively. The link functions $d_1, d_2: (0, \infty) \to \mathbb{R}$ are assumed to be strictly monotone and at least twice differentiable.
\end{definicao}

\begin{definicao}[Zero-adjusted Box-Cox symmetric regression models].
Let $\mathcal{Y}_1, \ldots, \mathcal{Y}_n$ be independent random variables where $\mathcal{Y}_i \sim \text{ZABCS}(\alpha_i, \mu_i, \sigma_i, \lambda; r)$ for $i=1,\ldots,n$. The ZABCS regression models are defined by the systematic components
\begin{equation*}
d_0(\alpha_i) = \bm{\mathcal{Z}}_i^{\top}\bm{\kappa}, \quad d_1(\mu_i) = \bm{\mathcal{X}}_i^{\top}\bm{\beta}, \quad \text{and} \quad d_2(\sigma_i) = \bm{\mathcal{S}}_i^{\top}\bm{\tau},
\end{equation*}
where $\bm{\kappa} \in \mathbb{R}^m$, $\bm{\beta} \in \mathbb{R}^p$, and $\bm{\tau} \in \mathbb{R}^q$ are unknown parameter vectors, alongside the unknown parameter $\lambda$, such that $m + p + q + 1 < n$. The terms $\bm{\mathcal{Z}}_i$, $\bm{\mathcal{X}}_i$, and $\bm{\mathcal{S}}_i$ are known covariate vectors corresponding to the full-rank model matrices $\bm{\mathcal{Z}} = [\bm{\mathcal{Z}}_1, \ldots, \bm{\mathcal{Z}}_n]^{\top}$, $\bm{\mathcal{X}} = [\bm{\mathcal{X}}_1, \ldots, \bm{\mathcal{X}}_n]^{\top}$, and $\bm{\mathcal{S}} = [\bm{\mathcal{S}}_1, \ldots, \bm{\mathcal{S}}_n]^{\top}$, respectively. The link functions $d_0: (0, 1) \to \mathbb{R}$ and $d_1, d_2: (0, \infty) \to \mathbb{R}$ are assumed to be strictly monotone and at least twice differentiable.
\end{definicao}

A standard choice for $d_1$ and $d_2$ is the logarithmic link function, which provides a straightforward interpretation of the regression coefficients as multiplicative effects; see \cite{demedeirosbcs}.  For the probability of exact zeros in the ZABCS framework, a natural choice for $d_0$ is the logit link.

Inference for both the BCS and ZABCS regression models is performed using the maximum likelihood method. 

For the BCS regression model, let $\bm{\theta} = (\bm{\beta}^\top, \bm{\tau}^\top, \lambda)^\top$ be the parameter vector. The maximum likelihood estimates are obtained by maximizing the log-likelihood function $\ell(\bm{\theta}) = \sum_{i=1}^n \ell_i(\mu_i, \sigma_i, \lambda)$, where
\begin{equation*}
\ell_i(\mu_i, \sigma_i, \lambda) = \begin{cases}
(\lambda - 1) \log y_i - \lambda \log \mu_i - \log  \sigma_i + \log r(z_i^2) - \log R\left(\dfrac{1}{\sigma_i |\lambda|}\right), & \mbox{ if } \lambda \neq 0,\\
-\log y_i - \log \sigma_i + \log r(z_i^2), & \mbox{ if } \lambda = 0,
\end{cases}
\end{equation*}
with $z_i = T(y_i; \mu_i, \sigma_i, \lambda)$. Because the corresponding score equations do not possess closed-form solutions, numerical optimization algorithms, such as quasi-Newton methods, are required. The optimization algorithms require the specification of an initial value for the iterative process, denoted by $\widetilde{\bm{\theta}} = (\widetilde{\bm{\beta}}^\top, \widetilde{\bm{\tau}}^\top, \widetilde{\lambda})^\top$. \cite{demedeirosbcs} recommend setting $\widetilde{\lambda} = 0$, $\widetilde{\bm{\beta}} = (\bm{X}^\top \bm{X})^{-1} \bm{X}^\top \bm{\upsilon}$, where $\bm{\upsilon} = (d_1(y_1), \ldots, d_1(y_n))^\top$. For $\widetilde{\bm{\tau}}$, we suggest $\widetilde{\bm{\tau}}= (\widetilde{\tau}, 0, \ldots, 0)^\top$, with $\widetilde{\tau} = \text{asinh}(\text{CV}_y/1.5)/\Phi^{-1}(0.75)$, where $\text{CV}_y = 0.75(Q_3-Q_1)/Q_2$, and $Q_1$, $Q_2$, and $Q_3$ denote the first, second and third sample quantile of $y$, respectively, $\Phi^{-1}(\cdot)$ is the quantile function of a random variable with standard normal distribution, and $\text{asinh}(\cdot)$ is the inverse hyperbolic sine function.

For the ZABCS regression models, the parameter vector is expanded to include the discrete component: $\bm{\theta} = (\bm{\kappa}^\top, \bm{\beta}^\top, \bm{\tau}^\top, \lambda)^\top$. A key analytical advantage of this zero-adjusted framework is that its log-likelihood function is perfectly separable into two independent components: $\ell^{(0)}(\bm{\theta}) = \ell_1(\bm{\kappa}) + \ell_2(\bm{\beta}, \bm{\tau}, \lambda)$,
where
$$
 \ell_1(\bm{\kappa}) = \sum_{i=1}^{n} \ell_i^{(0)}(\alpha_i) \quad \text{and} \quad
 \ell_2(\bm{\beta}, \bm{\tau}, \lambda) = \sum_{i \in \mathcal{I}}\ell_i (\mu_i,\sigma_i, \lambda),
$$
in which $\ell_i^{(0)}(\alpha_i)  = \ind(y_i = 0) \log (\alpha_i) + [1-\ind (y_i = 0)] \log (1-\alpha_i)$. Consequently, maximum likelihood estimation can be efficiently performed in two independent stages:

\begin{enumerate}
    \item Discrete component: The parameter $\bm{\kappa}$ is estimated by maximizing $\ell_1(\bm{\kappa})$. This is equivalent to fitting a standard generalized linear model for binary data \citep{mccullagh1989generalized}, using the indicator function $\mathbb{I}(y_i = 0)$ as the response and $d_0(\cdot)$ as the link function.
    
    \item Continuous component: The parameters $(\bm{\beta}^\top, \bm{\tau}^\top, \lambda)^\top$ are estimated by maximizing $\ell_2(\bm{\beta}, \bm{\tau}, \lambda)$. This step is mathematically equivalent to fitting a standard BCS regression model, but applied exclusively to the subset of strictly positive observations, $\mathcal{I} = \{i: y_i > 0\}$.
\end{enumerate}

The extra parameter $\zeta$, if any, is selected by minimizing the overall goodness-of-fit measure $\Upsilon_\zeta$, defined as
\begin{equation*}
\Upsilon_\zeta = n^{-1} \displaystyle \sum_{i=1}^n | \Phi^{-1}[F(y^{(i)}; \widehat{\mu_i}, \widehat{\sigma_i}, \widehat{\lambda})] - \upsilon^{(i)}|,
\end{equation*}
where $\widehat{\mu}_i = d_1^{-1}(\bm{x}_i^\top \widehat{\bm{\beta}})$, $\widehat{\sigma}_i = d_2^{-1}(\bm{z}_i^\top \widehat{\bm{\tau}})$, and $\widehat{\lambda}$ are the maximum likelihood estimates of $\mu_i$, $\sigma_i$, and $\lambda$, respectively, $y^{(i)}$ is the $i$th order statistic of the sample $y_1, \ldots, y_n$, $\upsilon^{(i)}$ is the mean of the $i$th order statistic in a random sample of size $n$ of the standard normal distribution, and $\Phi(\cdot)$ is the CDF of the standard normal distribution. Alternatively, $\zeta$ may be selected by maximizing $\ell(\widehat{\bm{\theta}}_\zeta)$, where $\widehat{\bm{\theta}}_\zeta$ is the maximum likelihood estimate of $\bm{\theta}$ for fixed $\zeta$.

Some diagnostic tools for the BCS and ZABCS regression models are presented in \cite{demedeirosbcs}, including quantile residuals, local influence methods, and goodness-of-fit measures. Applications and further details on inference methods are found in \cite{demedeirosbcs}. 

\section[R code]{R implementation} \label{sec:implementation}

The \CRANpkg{BCSreg} package offers a comprehensive regression modeling framework for non-negative responses based on the BCS and ZABCS classes of distributions, including inferential procedures and tools for assessing model fit. Parameter estimation and significance tests are based on likelihood theory. Residual analysis is implemented primarily using quantile residuals \citep{dunn1996}. However, for zero-adjusted models, Pearson residuals are also available for assessing the discrete component. Other diagnostic tools, such as local influence analysis and goodness-of-fit measures, are also available. The package also allows the log-symmetric and zero-adjusted log-symmetric regression model subclasses to be fitted by setting the parameter $\lambda$ to zero.

\subsection[R code]{BCS and ZABCS distributions in the \CRANpkg{BCSreg} package}\label{sec:BCS_ZABCS_R_implementation}

The implementation of the BCS and ZABCS distributions in the \CRANpkg{BCSreg} package uses the standard distribution framework in R. The functions \fct{dBCS}, \fct{pBCS}, \fct{qBCS}, and \fct{rBCS} compute the probability density function, cumulative distribution function, quantile function, and random samples of the BCS distributions. The syntax of these functions is:
\begin{Scode}
dBCS(x, mu, sigma, lambda, zeta, family = "NO", log = FALSE)
pBCS(q, mu, sigma, lambda, zeta, family = "NO", lower.tail = TRUE,
     log.p = FALSE)
qBCS(p, mu, sigma, lambda, zeta, family = "NO", lower.tail = TRUE,
     log.p = FALSE)
rBCS(n, mu, sigma, lambda, zeta, family = "NO")
\end{Scode}
As discussed in Section \ref{sec:BCSmodel}, the class of BCS distributions and its zero-adjusted extension are generated from truncated standard symmetric distributions, where the corresponding generating model is specified through the density generating function $r(\cdot)$. In the \CRANpkg{BCSreg} package, the generating distribution is specified through the \code{family} argument. Currently, the package includes eight generating distributions, namely, normal (\code{family = "NO"}), Student’s t (\code{family = "ST"}), power exponential (\code{family = "PE"}), type-I logistic  (\code{family = "LOI"}), type-II logistic (\code{family = "LOII"}), hyperbolic (\code{family = "HP"}), slash (\code{family = "SL"}), and sinh-normal (\code{family = "SN"}). The arguments \code{mu}, \code{sigma}, and \code{lambda} specify the parameters $\mu$, $\sigma$, and $\lambda$. If \code{lambda = 0}, these functions return results for the log-symmetric distributions. If the generating distribution depends on an additional parameter, its value must be specified in the \code{zeta} argument. On the other hand, if it does not depend on an extra parameter, the argument \code{zeta} is ignored. The arguments \code{x} and \code{q} are vectors of quantiles, \code{p} is a vector of probabilities, and \code{n} is the number of random numbers to be generated. Other arguments are \code{log}, \code{log.p}, and \code{lower.tail}. If \code{log = TRUE}, then the logarithm of the probability density function is returned. If \code{log.p = TRUE}, then the logarithm of the cumulative distribution function is returned and the quantile function is computed for $\exp(p)$. If \code{lower.tail = FALSE}, the upper tail of the cumulative distribution function is returned and the quantile function is computed for $1-p$. 

Analogous functions are available for the ZABCS class of probability models, namely \code{dZABCS()}, \code{pZABCS()}, \code{qZABCS()}, and \code{rZABCS()}. The syntax of these functions is:
\begin{Scode}
dZABCS(x, alpha, mu, sigma, lambda, zeta, family = "NO", log = FALSE)
pZABCS(q, alpha, mu, sigma, lambda, zeta, family = "NO", lower.tail = TRUE,
       log.p = FALSE)
qZABCS(p, alpha, mu, sigma, lambda, zeta, family = "NO", lower.tail = TRUE,
       log.p = FALSE)
rZABCS(n, alpha, mu, sigma, lambda, zeta, family = "NO")
\end{Scode}
These functions include the additional argument \code{alpha}, which specifies the probability of observing a zero response.

The density generating functions $r(\cdot)$ of all BCS (and ZABCS) distributions implemented in the \CRANpkg{BCSreg} package are listed below:
\begin{itemize}
\item Normal: $r(u) = (2\pi)^{-1/2}\exp(-u/2)$;
\item Student's $t$: $r(u) = \zeta^{\zeta/2}B(1/2,\zeta/2)^{-1}(\zeta + u)^{-(\zeta+1)/2}$, $\zeta>0$ and $B(\cdot,\cdot)$ is the beta function. When $\zeta = 1$, the Student's $t$ distribution reduces to the Cauchy distribution;
\item Power exponential: $r(u) = \zeta/[p(\zeta) 2^{1 + 1/\zeta} \Gamma(1/\zeta)] $ $ \exp\left\{-u^{\zeta/2}/(2 p(\zeta)^\zeta) \right\}$, $\zeta > 1$ and $p(\zeta) = 2^{-1/\zeta}\Gamma(1/\zeta)^{1/2}$. The power exponential distribution reduces to the Laplace distribution when $\zeta = 1$, and it coincides with the normal distribution when $\zeta = 2$;
\item Type-I logistic: $r(u) = c\; \e^{-u}(1 + \e^{-u})^{-2}$, where $c \approx 1.484300029$ is a normalizing constant, obtained from the relation $\int_{0}^\infty u^{-1/2} r (u) \dd u = 1$;
\item Type-II logistic: $r(u) = \exp\{ - u^{1/2} \}(1 + \exp\{ - u^{1/2} \})^{-2}$;
\item Hyperbolic: $r(u) = \exp\left\{ - \zeta \sqrt{1+u} \right\}/[2 K_1(\zeta)]$, where $K_s(\zeta) = \int_0^\infty \frac{x^{s-1}}{2}\times$
\\$ \exp\{ -\frac{\zeta}{2} \left(x + \frac{1}{x} \right) \} \dd x$ is the modified Bessel function of the second kind order and order $s$.
\item Slash: $r(u) =  \zeta  2^{\zeta } \Gamma(\zeta  + 0.5, 0.5u) / \left(\sqrt{\pi} u^{\zeta  + 0.5}\right)$, for $u>0$, and $r(u) = \zeta/[(\zeta+0.5)\sqrt{2\pi}]$, for $u=0$, where $\zeta>0$ and $\Gamma(a,x) = \int_0^x t^{a-1}e^{-t}dt$ is the lower incomplete gamma function. When $\zeta = 1$, the slash distribution coincides with the canonical slash distribution;
\item Sinh-normal: $r(u) = 1/(\zeta \sqrt{2 \pi}) \cosh(u^{1/2}) \exp\left[ - 2/\zeta^2 \sinh^2 (u^{1/2}) \right]$, where $\zeta>0$ and $\sinh(\cdot)$ and $\cosh(\cdot)$ denote the hyperbolic sine and cosine functions, respectively.
\end{itemize}

\begin{figure}[!ht]
\centering
\includegraphics[scale=0.58]{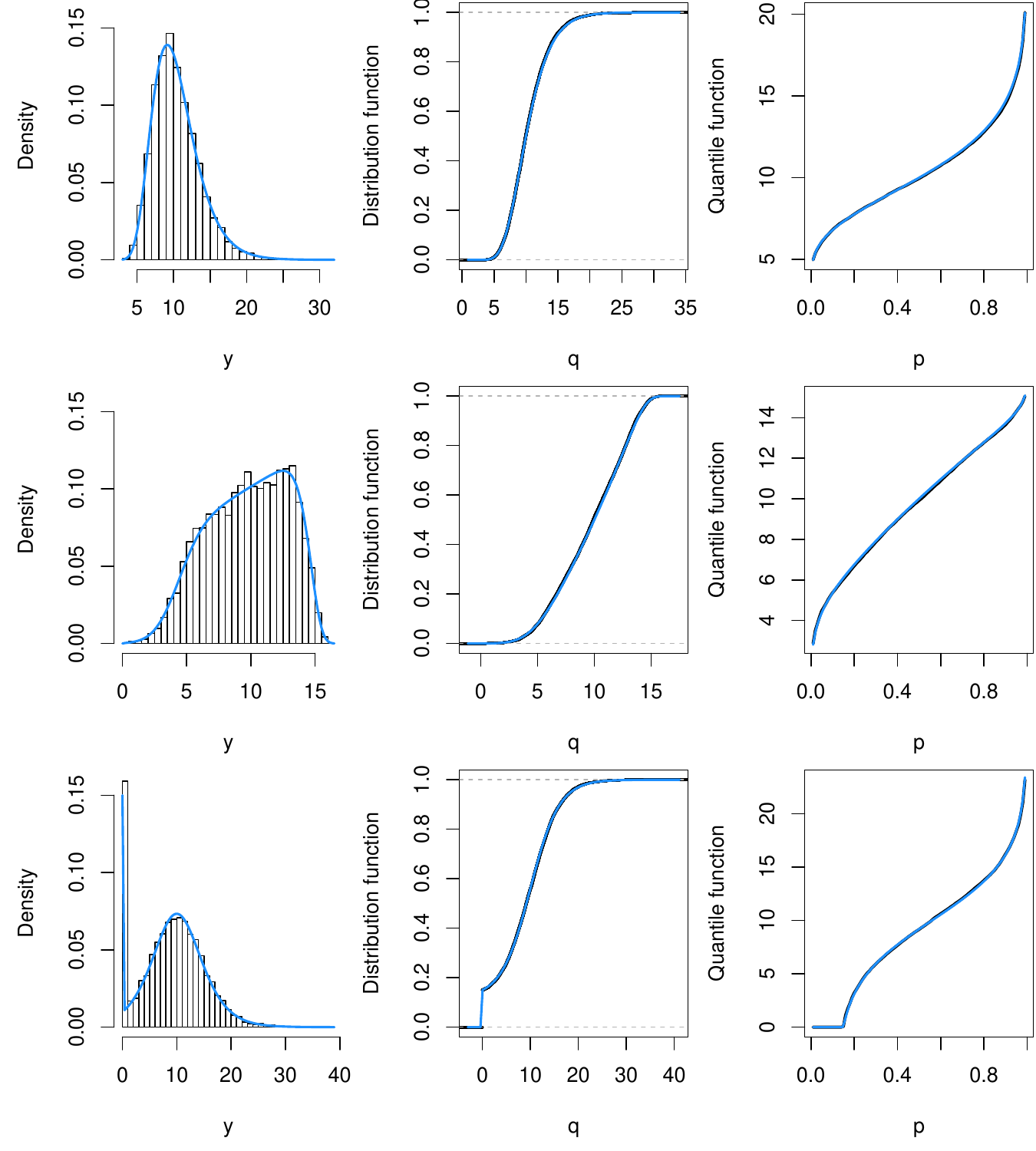}
\caption{\label{fig:BCSdist}
Histograms, empirical distribution functions, and sample quantile functions of random samples simulated from the log-normal, Box-Cox power exponential (with $\zeta = 6$), and zero-adjusted Box-Cox type-II logistic distributions. Solid lines represent the corresponding theoretical functions.}
\end{figure}

Figure~\ref{fig:BCSdist} illustrates the use of the distributional functions available in the \CRANpkg{BCSreg} package. It displays the histogram, empirical distribution function, and sample quantile function of random samples simulated from the log-normal, Box--Cox power exponential (with $\zeta = 6$), and zero-adjusted Box--Cox type-II logistic distributions. The parameters were fixed at $\mu = 10$ and $\sigma = 0.3$, while $\lambda$ was set to $0$, $1$, and $1.5$, respectively. The corresponding theoretical density, distribution, and quantile functions are superimposed on the plots.

\subsection{BCS and ZABCS regression models}

The \fct{BCSreg} function implements model fitting within the class of BCS and ZABCS regression models through a unified interface. The usage of the function is
\begin{Scode}
BCSreg(formula, data, subset, na.action, family = "NO", zeta, link = "log",   
       sigma.link = "log", alpha.link, control = BCSreg.control(...),
       model = FALSE, y = FALSE, x = FALSE, ...)
\end{Scode}

The \code{formula} argument is a symbolic description of the model, allowing the specification of different regression structures for the model parameters using the \CRANpkg{Formula} package \citep{zeileis2010}. As commonly implemented in regression modeling functions, the model specification provided through the \code{formula} argument consists of two main parts separated by ``\code{~}''. The left-hand side specifies the response variable, which must take values in either $(0,\infty)$ or $[0,\infty)$. The right-hand side can be further divided into up to three parts, separated by the ``\code{|}'' operator: (i) the first part specifies the model for the scale parameter $\mu$; (ii) the second part (optional) defines a regression structure for the relative dispersion parameter $\sigma$; and (iii) the third part (optional and only applicable for zero-adjusted data) defines the model for the zero-adjustment parameter $\alpha$. If only the first part is provided, the model includes a regression structure only for the scale parameter. If the third part is provided but the response variable contains no zeros, it is ignored. For instance, consider a dataset where \code{y} is the response variable, and \code{x}, \code{s}, and \code{z} are explanatory variables associated with the scale, relative dispersion, and zero-adjustment parameters, respectively. The following formulas illustrate different model specifications:
\begin{Scode}
formula = y ~ x         # Scale parameter only
formula = y ~ x | s     # Scale and relative dispersion parameters
formula = y ~ x | s | z # Scale, relative dispersion, and zero-adj. parameters
formula = y ~ x | 1 | z # Scale and zero-adj. parameters
formula = y ~ 1 | s | z # Relative dispersion and zero-adj. parameters  
\end{Scode}

The arguments \code{data}, \code{subset}, and \code{na.action} of the \fct{BCSreg} function specify the data set, an optional subset of observations, and the method for handling missing values, respectively. The argument \code{family} specifies the symmetric distribution used to generate the BCS or ZABCS model. As described in Section \ref{sec:BCS_ZABCS_R_implementation}, the currently supported families are \code{"NO"}, \code{"ST"}, \code{"PE"}, \code{"LOI"}, \code{"LOII"}, \code{"HP"}, \code{"SL"}, and \code{"SN"}. For the \code{"ST"}, \code{"PE"}, \code{"HP"}, \code{"SL"}, and \code{"SN"} families, the additional parameter $\zeta$ must be supplied through the \code{zeta} argument. 

The currently available link functions for the scale parameter $\mu$ and the relative dispersion parameter $\sigma$ are \code{"log"} (default), \code{"sqrt"}, \code{"inverse"},  and \code{"identity"}, which are specified through the \code{link} and \code{sigma.link} arguments, respectively. When fitting ZABCS models, the available link functions for the zero-adjustment parameter $\alpha$ are \code{"logit"} (default), \code{"probit"}, \code{"cloglog"}, \code{"cauchit"}, and \code{"loglog"}, which are specified through the \code{alpha.link} argument. The \code{control} argument specifies control options through the \fct{BCSreg.control} function, which can also be supplied directly to \fct{BCSreg} via \code{...}. Most of these options are passed directly to \fct{optim}, which is used to maximize the log-likelihood function. In addition to optimization settings, the user may provide initial values for the model parameters and optionally fix the skewness parameter $\lambda$ during estimation. A natural value to be fixed is \code{lambda = 0}, which corresponds to a log-symmetric regression model. Finally, the arguments \code{model}, \code{y}, and \code{x} are logical values indicating whether the corresponding components of the fitted model (respectively, the model frame, the response vector, and the model matrices) should be returned.

The \fct{BCSreg} function returns an \code{S3} object of class \class{BCSreg}, which contains several components associated with the fitted model, including the maximum likelihood estimates, fitted values, and the estimated asymptotic covariance matrix. A complete description of the object components is available in the package reference manual \citep{BCSreg}. 

Methods for extracting additional information from the fitted model are also available in the \CRANpkg{BCSreg} package. In addition to displaying a detailed summary of the fitted model, the \fct{summary} method returns an object of class \class{summary.BCSreg}, which contains summary tables with parameter estimates, standard errors, and individual significance tests for the regression coefficients, as well as residuals and goodness-of-fit measures. The residuals are computed using the method 
\begin{Scode}
residuals(object, approach = c("combined", "separated"), ...)
\end{Scode}
For a fitted BCS regression model, it returns the quantile residuals proposed by \citet{dunn1996}, which are defined by $r^q_i = \Phi^{-1}{F(y_i; \widehat{\mu}_i, \widehat{\sigma}_i, \widehat{\lambda})}, i = 1, \ldots, n,$ where $F$ denotes the CDF of the corresponding BCS distribution. For fitted ZABCS regression models, two approaches are available:
\begin{description}
\item[Combined approach:] Specified with \code{approach = "combined"} (default), it returns a single vector of residuals defined as
$$
r_i^q = \begin{cases}
\Phi^{-1}(u_i), & y_i = 0,\\
\Phi^{-1}{F^{(0)}(y_i; \widehat{\alpha}_i, \widehat{\mu}_i, \widehat{\sigma}_i, \widehat{\lambda})}, & y_i > 0,
\end{cases} \quad i = 1, \ldots, n,
$$
where $\widehat{\alpha}_i = d_0^{-1}(\bm{z}_i^\top \widehat{\bm{\kappa}})$ is the maximum likelihood estimate of $\alpha_i$, $u_i$ is a realization of a random variable uniformly distributed on $(0, \widehat{\alpha}_i]$, and $F^{(0)}$ denotes the CDF of the corresponding ZABCS distribution.

\item[Separated approach:] Specified with \code{approach = "separated"}, it returns a list containing the quantile residuals for the continuous component and the standardized Pearson residuals for the discrete component, defined by
$$
r_i^p = \dfrac{\mathbb{I}(y_i = 0) - \widehat{\alpha}_i}{\sqrt{\widehat{\alpha}_i(1-\widehat{\alpha}_i)(1-\widehat{h}_{ii})}}, \quad i = 1, \ldots, n,
$$
where $\widehat{h}_{ii}$ is the $i$th diagonal element of the hat matrix obtained by fitting a generalized linear model to the binary response $\mathbb{I}(y_i = 0)$; see \citet{demedeirosbcs} for details.
\end{description}

The \fct{plot} method provides seven types of plots for diagnostic analysis of a BCS or a ZABCS regression fit, including a plot of the residuals versus the fitted medians, a normal probability plot of the residuals with a confidence region constructed according to \citet{fox2015}, and an index plot of local influence based on the case-weight perturbation scheme. 

Another important method for objects of class \class{BCSreg} is the \fct{extra.parameter} function, which can be used to select the value of the extra parameter $\zeta$ in BCS and ZABCS models. As discussed in Section~\ref{sec:BCS_ZABCS_regression}, the value of $\zeta$ may be selected either by minimizing the upsilon goodness-of-fit statistic, $\Upsilon_\zeta$, or by maximizing the profile log-likelihood. The \fct{extra.parameter} function implements both approaches. Its usage is
\begin{Scode}
extra.parameter(object, family, grid = seq(1, 30, 2), trace = TRUE,
                plot = TRUE, control = BCSreg.control(...), ...)
\end{Scode}

The \code{family} argument specifies the generating family for which the extra parameter is to be selected. Note that the \code{"NO"}, \code{"LOI"}, and \code{"LOII"} families do not involve an extra parameter, and therefore this function is not applicable to these distributions. The \code{grid} argument is a numeric vector containing the candidate values of $\zeta$. The function produces plots of the upsilon statistic, $\Upsilon_\zeta$, and the profile log-likelihood as functions of $\zeta$, and returns an object of class \class{extra.parameter}. This object is a list containing the fitted models corresponding to each candidate value of $\zeta$ specified in \code{grid}. It also includes the associated log-likelihood and upsilon values for each fitted model. The \fct{extra.parameter} function requires an object of class \class{BCSreg} corresponding to a previously fitted model. Its implementation assumes that the Box-Cox normal (or zero-adjusted Box-Cox normal) distribution serves as the reference model within the BCS (or ZABCS) class and is therefore the first model to be fitted in an application. If necessary, this initial model can subsequently be refined by considering more flexible BCS (or ZABCS) distributions. Consequently, a natural choice for the \code{object} argument is the fitted Box-Cox normal (or zero-adjusted Box-Cox normal) regression model. Nevertheless, the initial fitted object is used only to extract the model specification, such as the regression structures, link functions, and other fitting options, before refitting the model over the candidate values of $\zeta$.

The list of all methods currently available for \class{BCSreg} objects is presented in Table~\ref{tab:methods}.
\begin{table}[ht!]
\centering
\begin{tabular}{lp{10cm}}
\hline
Function & Description  \\
\hline
\fct{print} &
Prints the estimated regression coefficients. \\

\fct{summary} &
Returns an object of class \class{summary.BCSreg}, containing detailed information about the fitted model, including parameter estimates, standard errors, residuals, and goodness-of-fit measures. The returned object also has a \fct{print} method that displays a detailed summary of the fitted model. \\

\fct{coef} &
Returns the estimated regression coefficients. The \code{model} argument specifies the submodel for which the coefficients are extracted and can be set to \code{"full"}, \code{"mu"}, \code{"sigma"}, or \code{"alpha"} (for zero-adjusted models). \\

\fct{vcov} &
Returns the estimated asymptotic covariance matrix of the regression coefficients. The \code{model} argument specifies the submodel of interest. \\

\fct{logLik} &
Returns the maximized log-likelihood. \\

\fct{model.matrix} &
Returns the design matrix associated with the selected regression submodel. \\

\fct{AIC} &
Returns the Akaike Information Criterion (AIC), Bayesian Information Criterion (BIC), or a generalized information criterion, depending on the value of \code{k}. \\

\fct{residuals} &
Returns quantile residuals for BCS regression models. For ZABCS regression models, the default (\code{approach = "combined"}) returns a single vector of combined quantile residuals, whereas \code{approach = "separated"} returns a list containing quantile residuals for the continuous component and standardized Pearson residuals for the discrete component. \\

\fct{plot} &
Produces diagnostic plots. Currently, seven types of plots are available: residuals versus fitted values, residual index plot, density plot, normal probability plot,     local influence plot based on the case-weight perturbation scheme, fitted versus observed values, and the $v(z)$ function versus the residuals. \\

\fct{influence} &
Computes influence measures for BCS and ZABCS regression models. \\

\fct{envelope} &
Produces a normal probability plot with simulated envelopes for the residuals. \\

\fct{extra.parameter} &
Provides graphical tools for selecting the extra parameter $\zeta$, when applicable. \\
\hline
\end{tabular}
\caption{\label{tab:methods} Methods available for objects of class \class{BCSreg}.}
\end{table}

\section{Examples using the \CRANpkg{BCSreg} package} \label{sec:illustrations}

To demonstrate the practical capabilities of the \CRANpkg{BCSreg} package, we present several illustrative examples. The applications utilize three real-world datasets included in the package: \code{raycatch}, \code{renewables2015}, and \code{education}. All analyses and computations were conducted using R version 4.6.1.

\subsection[R code]{\code{raycatch} data: IID setting}\label{raycatchiid}

As an initial illustration of the \CRANpkg{BCSreg} package, we consider the \code{raycatch} dataset, which contains information on artisanal white ray landings originally collected and studied by \cite{marion2015}. The data comprise 186 observations obtained from fishing trips between January 2012 and January 2013 that employed the traditional ``grozeira'' (a type of bottom longline) as gear in Baía de Todos os Santos, Bahia, Brazil. In this first scenario, our primary interest lies in modeling the catch per unit effort (\code{cpue}) under an independent and identically distributed (IID) framework. The response variable represents the fishing productivity (in grams) divided by the product of the number of hooks and the immersion time (in hours). The dataset can be loaded into the R session by:
\begin{Schunk}
\begin{Sinput}
R> data("raycatch", package = "BCSreg")
\end{Sinput}
\end{Schunk}
A preliminary visual analysis of the response variable, including its histogram and boxplot, is displayed in Figure~\ref{fig:ex1-fig112}. Some summary measures of \code{cpue} are also presented below. Notably, the empirical distribution of \code{cpue} exhibits a pronounced right skewness, reflected by a sample mean ($18.007$) that is substantially higher than the median ($11.127$), with observations ranging from a minimum of $0.596$ to a maximum of $166.667$. 
\begin{figure}[!h]
\centering
\includegraphics[scale=0.4]{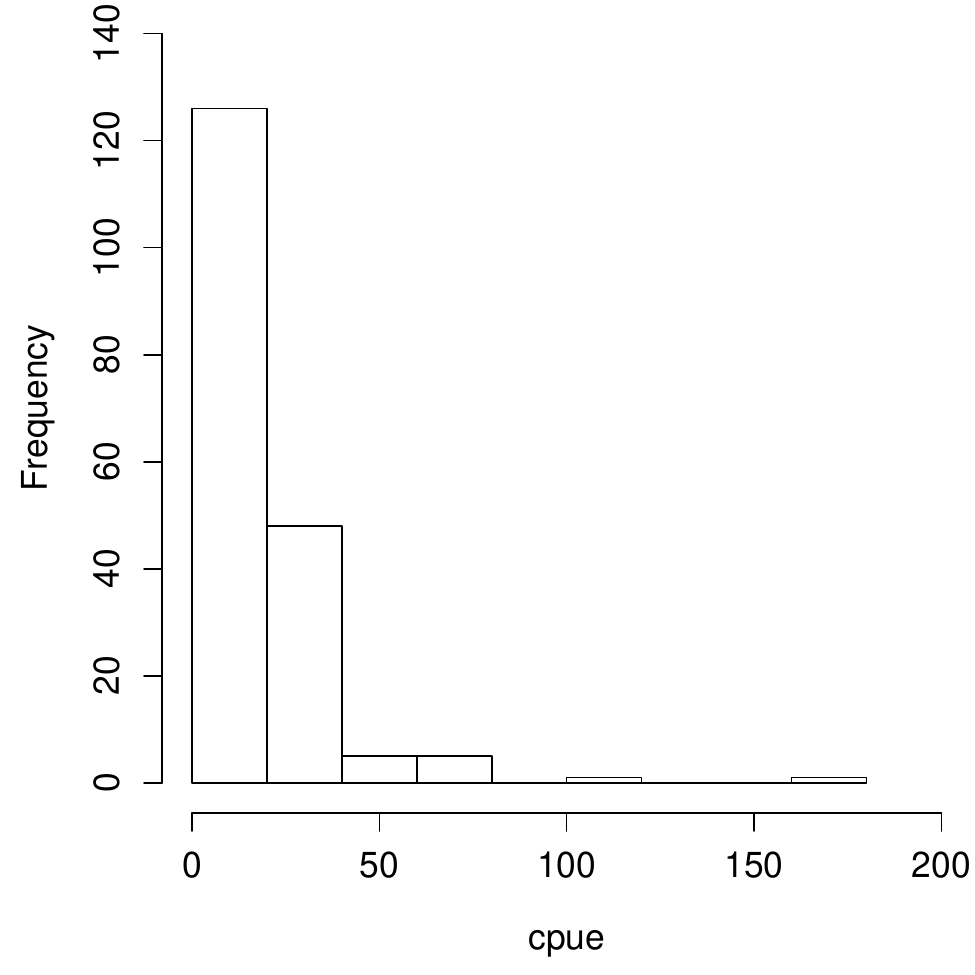}
\includegraphics[scale=0.4]{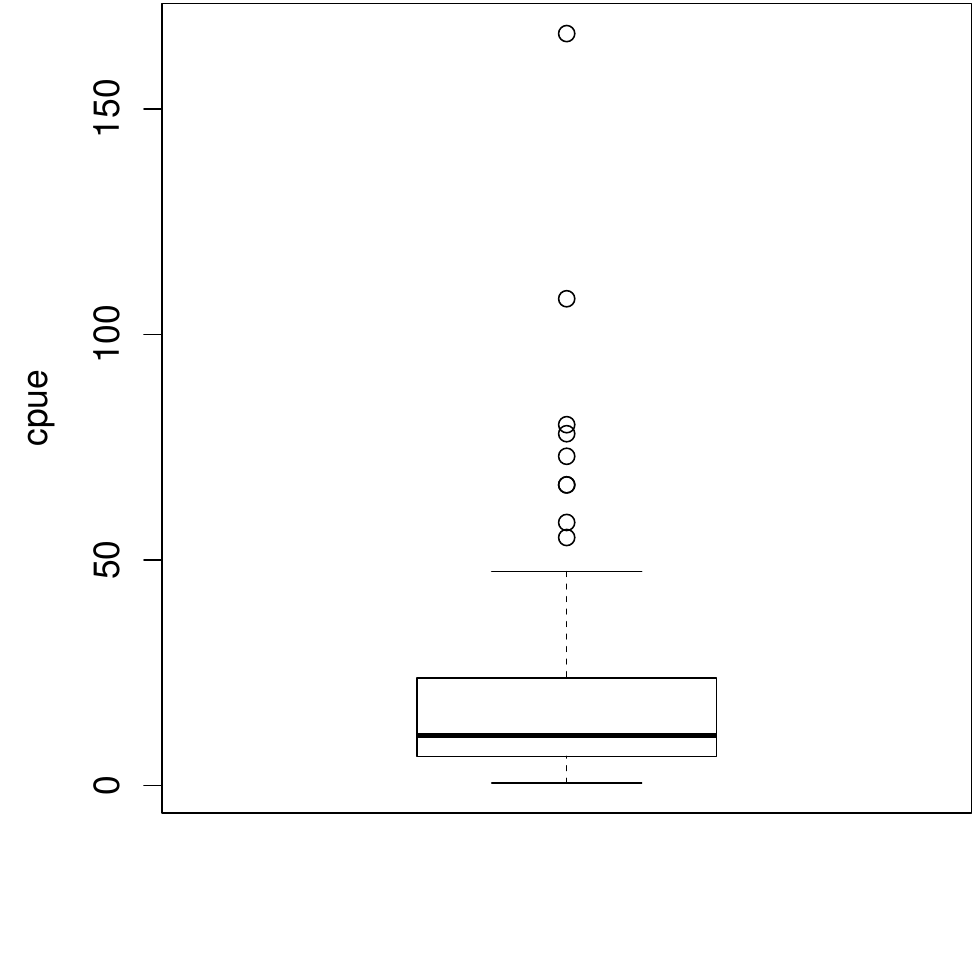}
\caption{\label{fig:ex1-fig112} Histogram (left side) and boxplot (right side) of the response variable -- \code{raycatch} data.}
\end{figure}

\begin{Schunk}
\begin{Sinput}
R> summary(raycatch$cpue)
\end{Sinput}
\begin{Soutput}
   Min. 1st Qu.  Median    Mean 3rd Qu.    Max. 
  0.596   6.542  11.127  18.007  23.794 166.667 
\end{Soutput}
\end{Schunk}

We now fit the \code{cpue} variable using the BCNO and BCSL distributions. For the BCSL distribution, we use the \fct{extra.parameter} function to select an optimum value of $\zeta$. In the \CRANpkg{BCSreg} package, it can be done via

\begin{Schunk}
\begin{Sinput}
R> BCNO_rc <- BCSreg(cpue ~ 1, data = raycatch, family = "NO")
R> BCSL_rc_zeta <- extra.parameter(BCNO_rc, family = "SL", grid = seq(0.5, 10, 0.5))
\end{Sinput}
\begin{Soutput}
Best value for zeta according to Upsilon: 2.5 and Profile log-lik.: 5 
\end{Soutput}
\begin{Sinput}
R> BCSL_rc <- BCSL_rc_zeta[[which.min(BCSL_rc_zeta$Upsilon)]]
\end{Sinput}
\end{Schunk}

\begin{figure}[!h]
\centering
\includegraphics[scale=0.4]{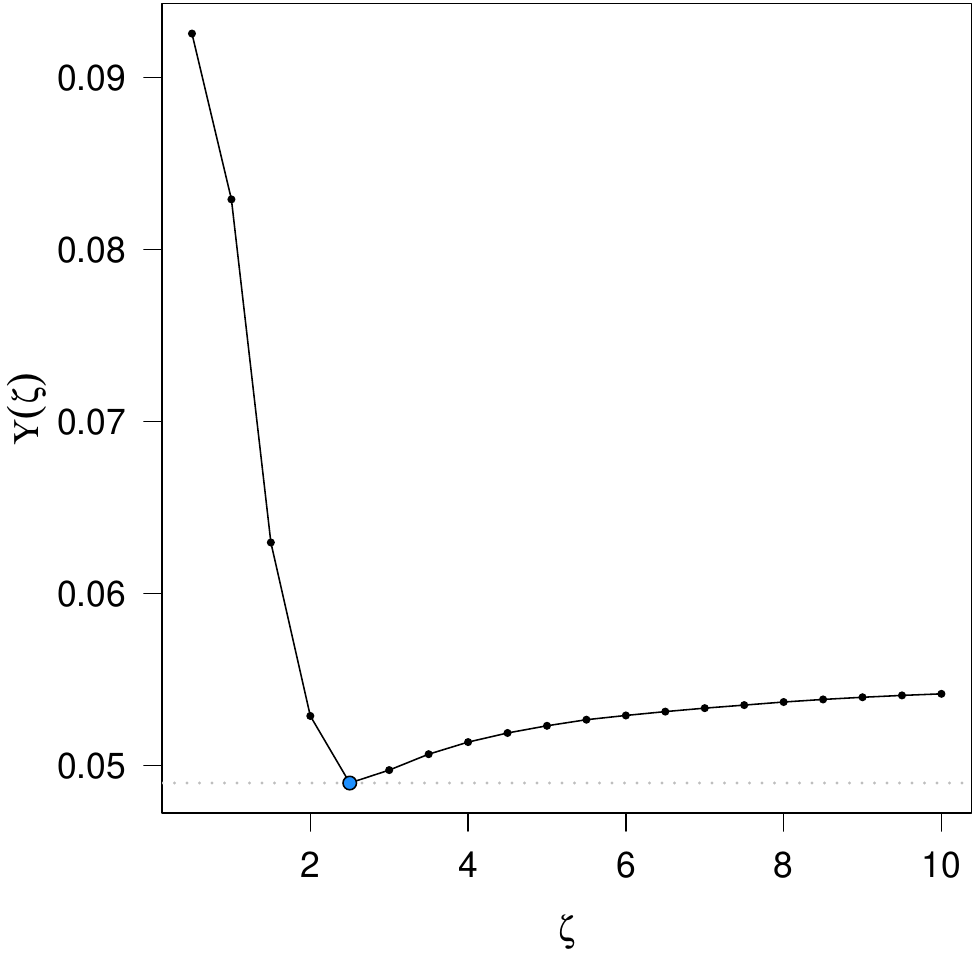}
\includegraphics[scale=0.4]{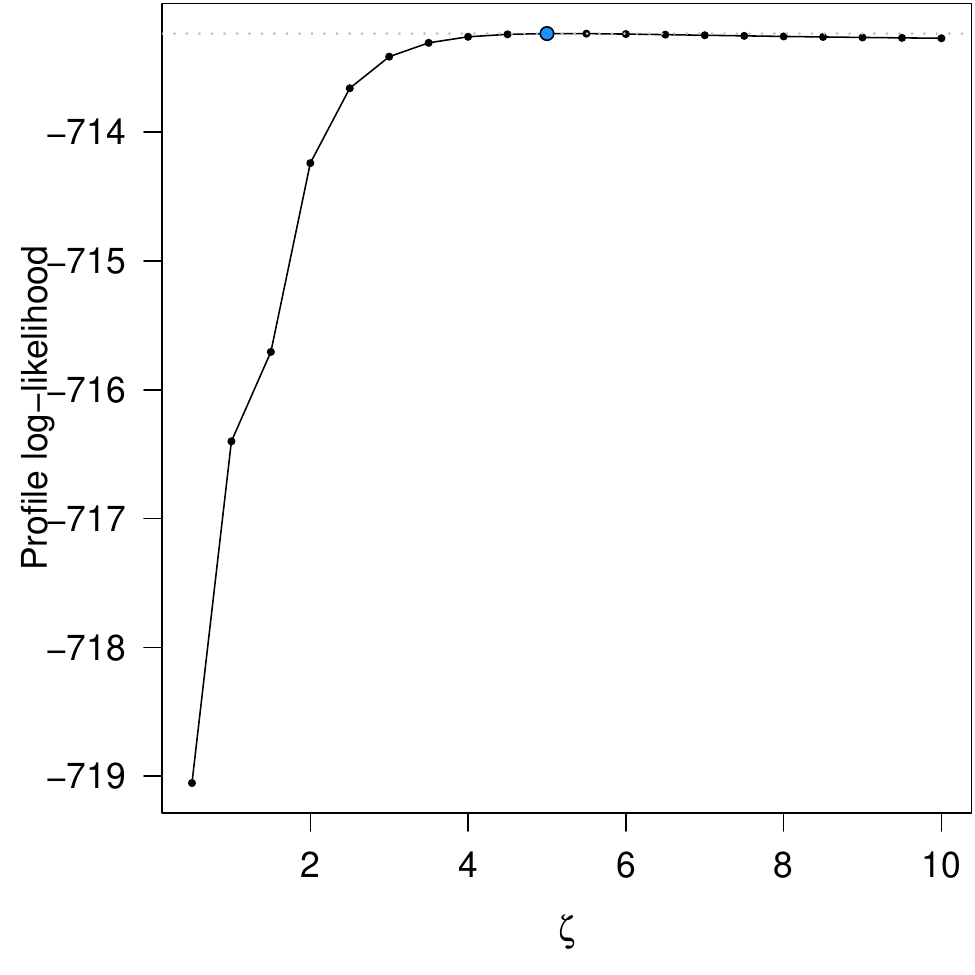}
\caption{\label{fig:ex1-fig3} Plot returned by the \fct{extra.parameter} for selecting an optimum value of $\zeta$ of the fit of the BCSL distribution -- \code{raycatch} data.}
\end{figure}

The \fct{extra.parameter} returns the optimum values of $\zeta$ based on two measures and present the plots of this measures as functions of $\zeta$; see Figure~\ref{fig:ex1-fig3}. We choose \code{zeta = 2.5}. 

To select between the BCNO and BCSL distributions, we compute the $\Upsilon_\zeta$ and the AIC for both fits. 
\begin{Schunk}
\begin{Sinput}
R> BCNO_rc_measures <- c(summary(BCNO_rc)$Upsilon.zeta, AIC(BCNO_rc))
R> BCSL_rc_measures <- c(summary(BCSL_rc)$Upsilon.zeta, AIC(BCSL_rc))
R> measures <- round(rbind(BCNO_rc_measures, BCSL_rc_measures),3)
R> colnames(measures) <- c("Upsilon", "AIC")
R> measures
\end{Sinput}
\begin{Soutput}
                 Upsilon      AIC
BCNO_rc_measures   0.055 1432.613
BCSL_rc_measures   0.049 1435.319
\end{Soutput}
\end{Schunk}

Since the values of $\Upsilon_\zeta$ and AIC for both fits are similar, we select the BCNO distribution, which offers a more parsimonious framework with fewer parameters. The \code{summary} output of the BCNO fit is presented below.

\begin{Schunk}
\begin{Sinput}
R> summary(BCNO_rc)
\end{Sinput}
\begin{Soutput}
Call:
BCSreg(formula = cpue ~ 1, data = raycatch, family = "NO")

Quantile residuals:
       Min         1Q     Median         3Q        Max 
-2.7602276 -0.6481854 -0.1073523  0.7181256  3.1414665 

Scale submodel with log link:
              Estimate Std. error  z value   Pr(>|z|)    
(Intercept) 2.51154614 0.07464441 33.64681 < 2.22e-16 ***

Relative dispersion submodel with log link:
               Estimate  Std. error  z value Pr(>|z|)
(Intercept) -0.05455300  0.05259796 -1.03717  0.29966

Skewness parameter:
           Estimate Std. error z value Pr(>|z|)
(lambda) 0.09990230 0.06085135 1.64174  0.10064
---
Signif. codes:  0 '***' 0.001 '**' 0.01 '*' 0.05 '.' 0.1 ' ' 1 

Generating family: NO (Box-Cox normal)
Log-likelihood: -713.3065 on 3 Df
Upsilon statistic: 0.05502085
AIC: 1432.613 and BIC: 1984.613
Number of iterations in optimization: 10
\end{Soutput}
\end{Schunk}

The estimated scale parameter based on the fit is $\widehat{\mu} = \exp(2.51155) \approx 12.32$, which is close to the sample median of $11.127$.

To assess the adequacy of the distributional assumptions of the fitted model, the \CRANpkg{BCSreg} package provides the \fct{envelope} function. This function constructs a normal probability plot of the quantile residuals enhanced with simulated envelopes. If the model is correctly specified, the observed residuals are expected to fall randomly within these boundaries. Figure~\ref{fig:Fig4} presents the normal probability plot of the quantile residuals with a simulated envelope for the \code{BCNO\_rc} fit and the histogram of \code{cpue} overlaid with the estimated density. These plots can be obtained as follows:
\begin{figure}[!h]
\centering
\includegraphics[scale=0.4]{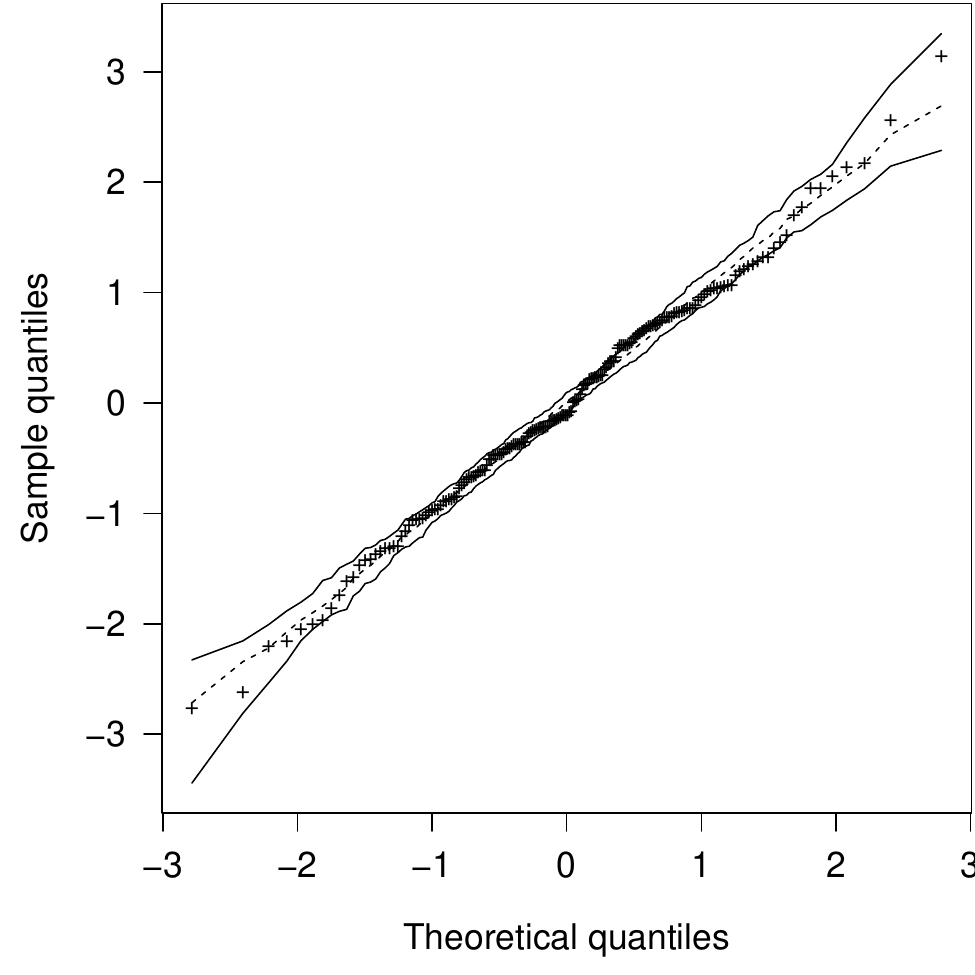}
\includegraphics[scale=0.4]{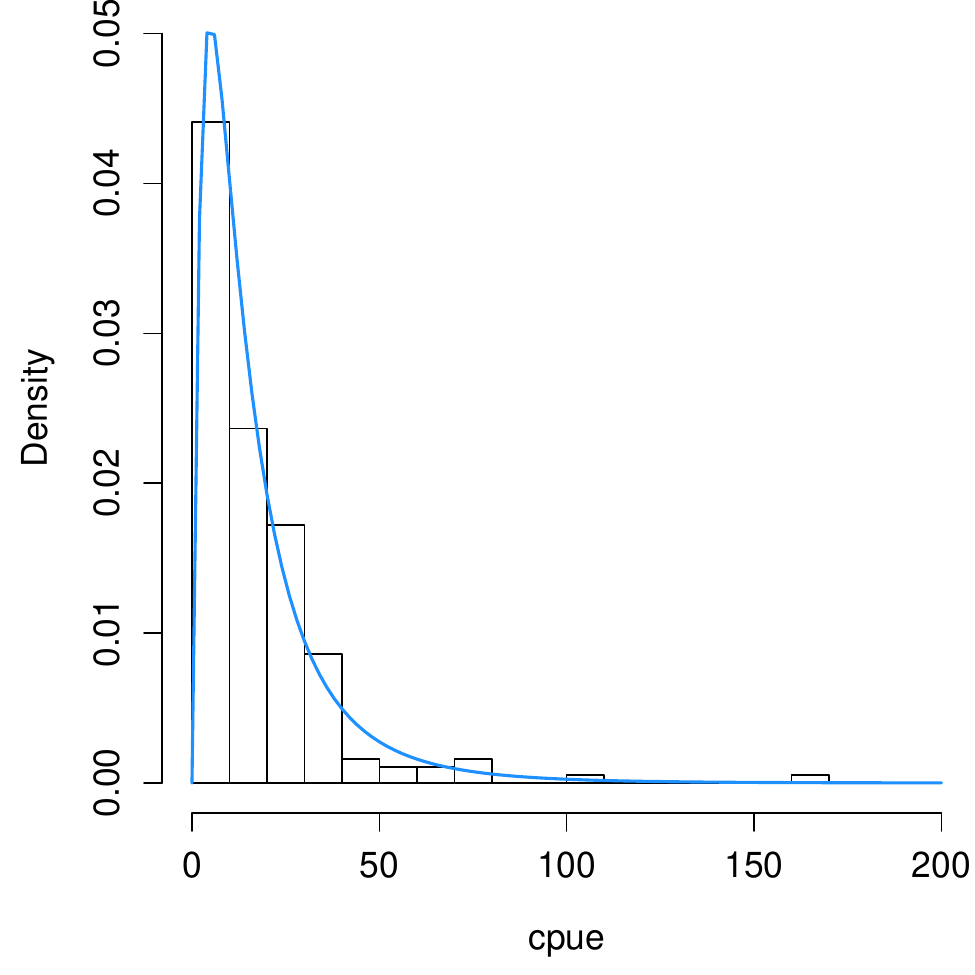}
\caption{\label{fig:Fig4} Normal probability plot of the quantile residual with simulated envelope for the \code{BCNO\_rc} fit and the histogram of \code{cpue} with the estimated density -- \code{raycatch} data.}
\end{figure}
\begin{Schunk}
\begin{Sinput}
R> envelope(BCNO_rc)
R> hist(raycatch$cpue, main = " ", xlab = "cpue", prob = TRUE, xlim = c(0, 200),
+      nclass = 20, ylim = c(0, 0.05), col = "white")
> curve(dBCS(x, BCNO_rc$fitted.values[1], exp(BCNO_rc$coefficients$sigma), 
+            BCNO_rc$lambda, family = "NO"), 0 , 200, add = TRUE, lwd = 2, 
+       col= "dodgerblue")
\end{Sinput}
\end{Schunk}

Since $\widehat{\sigma}\widehat{\lambda} = 0.094 \approx 0$, the estimate $\widehat{\mu}$ can be interpreted as the estimated median of \texttt{cpue}.
Using the delta method, an approximated 95\% confidence interval for the median \texttt{cpue} is 
\[
\left[\widehat{\mu} \mp 1.96 \times \mbox{se}(\widehat{\beta}) \exp(\widehat{\beta})\right] = [10.521,~ 14.127].
\]

\subsection[R code]{\code{renew\_elec\_output} data: ZABCS regression model}

To illustrate the performance of the \CRANpkg{BCSreg} package in a multiple regression framework, we analyze the \code{renewables2015} dataset. This dataset provides macroeconomic, environmental, and institutional indicators for $186$ countries in the year 2015, compiled from the World Bank database. The data can be loaded using:
\begin{Schunk}
\begin{Sinput}
R> data("renewables2015", package = "BCSreg")
\end{Sinput}
\end{Schunk}
The absolute renewable electricity output (\code{renew\_elec\_output}, measured in TWh) is heavily influenced by cross-country scale effects, as geographically massive nations naturally generate higher volumes of electricity. To mitigate this size bias and focus on renewable energy intensity, we construct a relative response variable, denoted as \code{y}. This variable is defined as the renewable electricity output divided by the actual agricultural land area (obtained by reverting the logarithm of \code{agri\_land}):
\begin{equation*}
\texttt{y}_i = \frac{\texttt{renew\_elec\_output}_i}{\exp(\texttt{agri\_land}_i)}.
\end{equation*}
\begin{Schunk}
\begin{Sinput}
R> y <- renewables2015$renew_elec_output/exp(renewables2015$agri_land)
R> renewables2015$y <- y
\end{Sinput}
\end{Schunk}
This response variable is semi-continuous on the non-negative real line, containing exact zeros. Specifically, 22 out of the 186 observations are exactly zero, which naturally motivates the use of the zero-adjusted BCS regression models. The primary objective is to investigate how institutional quality (\code{gov\_effec}), education expenditures (\code{adj\_sav\_edu}), and reliance on non-renewable energy sources (\code{elec\_fossil}) influence this renewable intensity. We specify a comprehensive structure where these three explanatory variables are included simultaneously across all three submodels (conditional scale, conditional relative dispersion, and zero-adjusted parameters). We specify this model as follows.
\begin{Schunk}
\begin{Sinput}
R> form_renew <- y ~ elec_fossil + gov_effec + adj_sav_edu | 
+   elec_fossil + gov_effec + adj_sav_edu | 
+   elec_fossil + gov_effec + adj_sav_edu 
R> ZABCNO_renew   <- BCSreg(form_renew, data = renewables2015, family = "NO") 
R> #extra.parameter(ZABCNO_renew, "ST", grid = seq(1, 10, 0.5)) # zeta = 2
R> #extra.parameter(ZABCNO_renew, "PE", grid = seq(0.5, 2, 0.1)) # zeta = 1.5 
R> ZABCT_renew    <- BCSreg(form_renew, data = renewables2015, family = "ST", 
+                        zeta = 2)
R> ZABCPE_renew   <- BCSreg(form_renew, data = renewables2015, family = "PE", 
+                        zeta = 1.5)
R> ZABCLOI_renew  <- BCSreg(form_renew, data = renewables2015, family = "LOI")
R> ZABCLOII_renew <- BCSreg(form_renew, data = renewables2015, family = "LOII")
\end{Sinput}
\end{Schunk}
Note that we used the \fct{extra.parameter} function to select the optimum value for $\zeta$ in the ZABCT and ZABCPE models.

To compare these models and select the most appropriate one, we compute the $\Upsilon_\zeta$ measure and the AIC for all fits.
\begin{Schunk}
\begin{Sinput}
R> modelos_renew <- list(
+   ZABCNO   = ZABCNO_renew,
+   ZABCT    = ZABCT_renew,
+   ZABCPE   = ZABCPE_renew,
+   ZABCLOI  = ZABCLOI_renew,
+   ZABCLOII = ZABCLOII_renew
+ )
R> measures <- round(t(sapply(modelos_renew, function(m) c(summary(m)$Upsilon.zeta,
+                                                         AIC(m)))), 3)
R> colnames(measures) <- c("Upsilon", "AIC")
R> measures
\end{Sinput}
\begin{Soutput}
         Upsilon       AIC
ZABCNO     0.076 -2465.660
ZABCT      0.049 -2462.127
ZABCPE     0.053 -2466.986
ZABCLOI    0.149 -2449.119
ZABCLOII   0.053 -2469.588
\end{Soutput}
\end{Schunk}

Note that the ZABCLOI model seems unsuitable for fitting the data. While the ZABCLOII model yields the lowest AIC, the ZABCT model presents the lowest $\Upsilon_\zeta$ measure ($\Upsilon_\zeta = 0.049$) and a highly competitive AIC. Thus, we further investigate the ZABCT fit. The \code{summary} output of the ZABCT fit is presented below.

\begin{Schunk}
\begin{Sinput}
R> summary(ZABCT_renew)
\end{Sinput}
\begin{Soutput}
Call:
BCSreg(formula = form_renew, data = renewables2015, family = "ST", zeta = 2)

Quantile residuals:
       Min         1Q     Median         3Q        Max 
-2.6962086 -0.5916737  0.0195977  0.6268790  2.7859632 

--- Fit for the discrete component ---

Zero-adjustment submodel with logit link:
                Estimate   Std. error  z value  Pr(>|z|)   
(Intercept) -26.90042294   9.00329450 -2.98784 0.0028095 **
elec_fossil   0.27640964   0.09213975  2.99990 0.0027007 **
gov_effec    -0.37515365   0.33582631 -1.11711 0.2639489   
adj_sav_edu  -0.18602393   0.14302432 -1.30065 0.1933798   

--- Fit for the continuous component ---

Scale submodel with log link:
                Estimate   Std. error   z value   Pr(>|z|)    
(Intercept) -7.924142490  0.494510546 -16.02421 < 2.22e-16 ***
elec_fossil -0.017138090  0.004449473  -3.85171 0.00011729 ***
gov_effec    1.529195157  0.153462130   9.96464 < 2.22e-16 ***
adj_sav_edu -0.174679737  0.105796663  -1.65109 0.09872034 .  

Relative dispersion submodel with log link:
                Estimate   Std. error  z value Pr(>|z|)
(Intercept) -0.282567216  0.322298807 -0.87672  0.38064
elec_fossil  0.004535112  0.003099890  1.46299  0.14347
gov_effec   -0.157704836  0.108224426 -1.45720  0.14506
adj_sav_edu  0.072905229  0.062212534  1.17187  0.24125

Skewness parameter:
          Estimate Std. error z value Pr(>|z|)   
(lambda) 0.3775001  0.1430243 2.63941 0.008305 **
---
Signif. codes:  0 '***' 0.001 '**' 0.01 '*' 0.05 '.' 0.1 ' ' 1 

Generating family: ST(2) (Box-Cox t with zeta = 2)
Log-likelihood: 1245.063 on 14 Df
Upsilon statistic: 0.04890049
AIC: -2462.127 and BIC: 113.8731
Number of iterations in optimization: 24
\end{Soutput}
\end{Schunk}

The output of the \code{summary} method provides a comprehensive overview of the fitted ZABCT model. In the discrete component (zero-adjustment submodel), the \code{elec\_fossil} covariate is statistically significant at the $1\%$ level, indicating its influence on the probability of observing a strictly zero renewable intensity. For the continuous component, both \code{elec\_fossil} and \code{gov\_effec} are highly significant in the scale submodel, presenting negative and positive effects, respectively. Furthermore, none of the covariates seem to significantly affect the relative dispersion. Finally, the statistical significance of the skewness parameter ($\lambda$) highlights the asymmetry of the positive observations, emphasizing the importance of adopting a flexible distribution such as the zero-adjusted Box-Cox $t$ to properly model the data. 

Based on the significance levels observed in the full model, a more parsimonious specification can be pursued by excluding non-significant covariates (using a $5\%$ significance level) and treating the relative dispersion as constant. Given that the \CRANpkg{BCSreg} formula framework structures the components as \code{scale | relative dispersion | zero-adjusted}, the \code{adj\_sav\_edu} variable is removed from the first part (scale submodel), the second part (relative dispersion) is set to a constant (\code{1}), and both \code{gov\_effec} and \code{adj\_sav\_edu} are dropped from the third part (zero-adjustment submodel). The updated, reduced model formula is defined and fitted as follows:
\begin{Schunk}
\begin{Sinput}
R> form_reduced <- y ~ elec_fossil + gov_effec | 1 | elec_fossil
R> ZABCT_reduced <- BCSreg(form_reduced, data = renewables2015, 
+                          family = "ST", zeta = 2)
R> summary(ZABCT_reduced)
\end{Sinput}
\begin{Soutput}
Call:
BCSreg(formula = form_reduced, data = renewables2015, family = "ST", zeta = 2)

Quantile residuals:
       Min         1Q     Median         3Q        Max 
-3.0212677 -0.6178534  0.0436094  0.5846922  2.8703960 

--- Fit for the discrete component ---

Zero-adjustment submodel with logit link:
                Estimate   Std. error  z value  Pr(>|z|)   
(Intercept) -27.87224328   8.90981555 -3.12826 0.0017584 **
elec_fossil   0.27943909   0.09136102  3.05862 0.0022236 **

--- Fit for the continuous component ---

Scale submodel with log link:
                Estimate   Std. error   z value   Pr(>|z|)    
(Intercept) -8.927540754  0.399179313 -22.36474 < 2.22e-16 ***
elec_fossil -0.015124384  0.003757249  -4.02539 5.6881e-05 ***
gov_effec    1.398566148  0.129283039  10.81786 < 2.22e-16 ***

Relative dispersion submodel with log link:
             Estimate Std. error z value Pr(>|z|)
(Intercept) 0.4679484  0.3334322 1.40343  0.16049

Skewness parameter:
           Estimate Std. error z value   Pr(>|z|)    
(lambda) 0.45936484 0.09136102 5.02802 4.9558e-07 ***
---
Signif. codes:  0 '***' 0.001 '**' 0.01 '*' 0.05 '.' 0.1 ' ' 1 

Generating family: ST(2) (Box-Cox t with zeta = 2)
Log-likelihood: 1238.362 on 8 Df
Upsilon statistic: 0.07013821
AIC: -2460.724 and BIC: -988.724
Number of iterations in optimization: 22
\end{Soutput}
\end{Schunk}
The summary of the parsimonious model reveals that all remaining covariates, as well as the skewness parameter ($\lambda$), are now highly statistically significant (all with $p$-values $< 0.01$). When comparing this reduced specification with the full model, we observe a substantial gain in parsimony, reducing the number of estimated parameters from $14$ to $8$ degrees of freedom. Although the AIC increased slightly from $-2462.127$ to $-2460.724$, this difference is strictly negligible, meaning the loss in goodness-of-fit is well compensated by the simplicity of the model.

\begin{figure}[!h]
\centering
\includegraphics[scale=0.4]{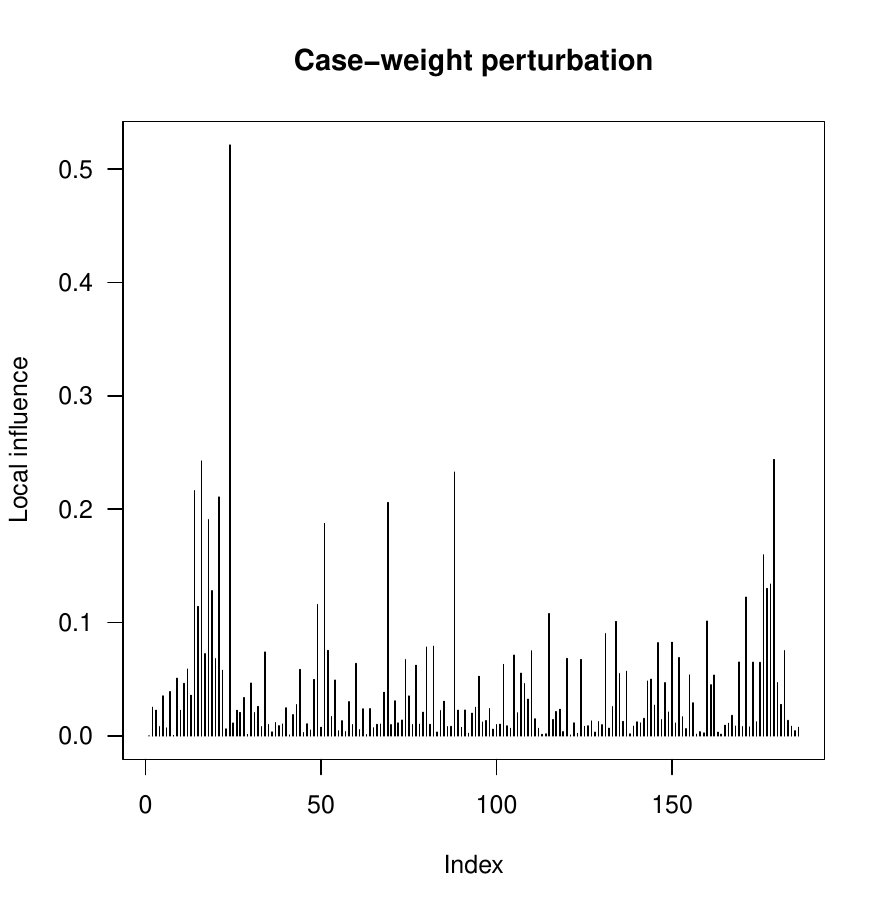}
\includegraphics[scale=0.4]{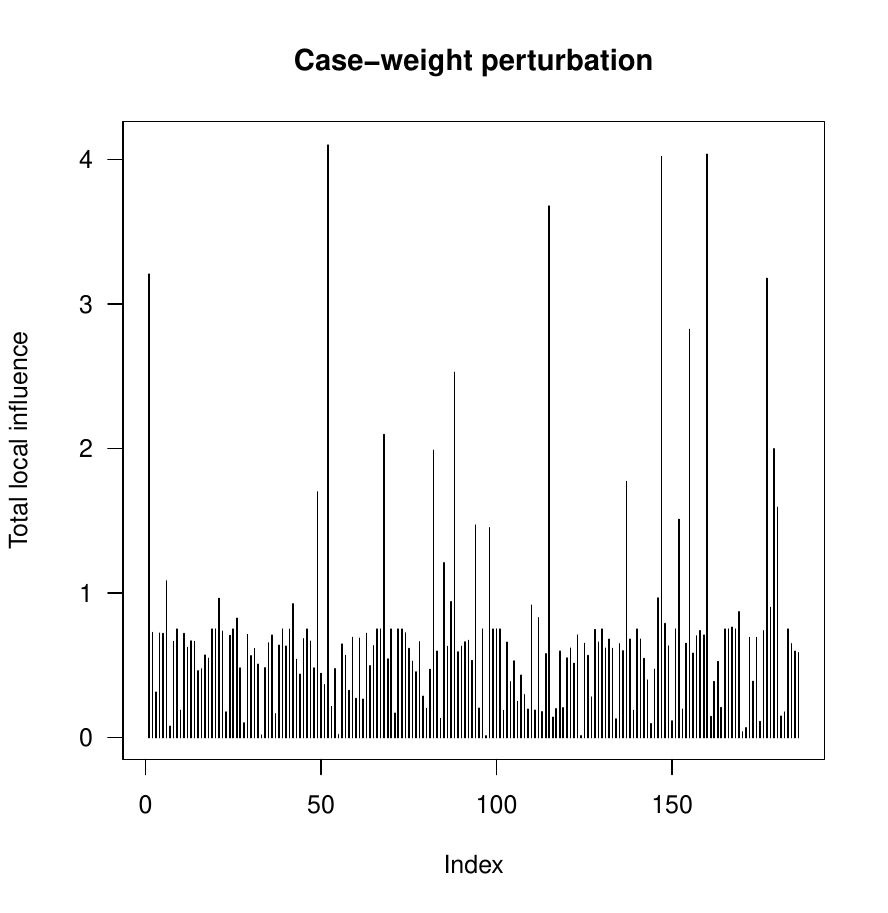}
\caption{\label{fig:Fig5} Influence plots for the for the \code{ZABCT\_reduced} fit -- \code{renew\_elec\_output} data.}
\end{figure}

Influence diagnostics can be performed in \CRANpkg{BCSreg} package through the \fct{influence} function. The results are presented in Figure \ref{fig:Fig5}.
\begin{Schunk}
\begin{Sinput}
R> inf_measures <- influence(ZABCT_reduced)
R> plot(inf_measures)
\end{Sinput}
\end{Schunk}

\subsection[R code]{\code{education} data: fixing the skewness parameter}

To further demonstrate the flexibility of the \CRANpkg{BCSreg} package, we consider the \code{education} dataset, which contains household-level data on expenditures on basic education in the state of São Paulo, Brazil, derived from the 2017–2018 Brazilian Consumer Expenditure Survey (POF/IBGE). This dataset can be loaded via:
\begin{Schunk}
\begin{Sinput}
R> data("education", package = "BCSreg")
\end{Sinput}
\end{Schunk}
Unlike the previous example, this microdata level application features a large sample size ($4,232$ households) and an extreme case of zero-inflation: approximately $93\%$ of the households reported exactly zero expenditure on basic education over the 12-month period. 

Our primary goal here is to investigate the effect of the reference person's attributes—specifically age (\code{age}), education (\code{years\_schooling}), and per capita income (\code{income\_pc})—across all submodels. We aim to illustrate a feature of the \CRANpkg{BCSreg} package: the ability to fit a ZABCS regression model with fixed $\lambda$. Thus, we can compare, for instance, the ZABCS framework against the standard zero-adjusted log-symmetric models ($\lambda = 0$).

We specify the three-part formula and fit both the unrestricted BCS model (estimating $\lambda$) and the restricted log-symmetric model (fixing $\lambda = 0$) using the \code{"LOII"} family:
\begin{Schunk}
\begin{Sinput}
R> form_educ <- expense ~ age + years_schooling + income_pc | 
+   age + years_schooling + income_pc | 
+   age + years_schooling + income_pc
R> ZABCNO_educ <- BCSreg(form_educ, data = education, family = "LOII")
R> ZALS_educ <- BCSreg(form_educ, data = education, family = "LOII",
+                     control = BCSreg.control(lambda = 0))
\end{Sinput}
\end{Schunk}
We can construct a 95\% asymptotic Wald confidence interval to test the null hypothesis $\mathcal{H}_0: \lambda = 0$ against the two-sided alternative $\mathcal{H}_1: \lambda \neq 0$. In R, this can be performed by extracting the estimate and its corresponding standard error directly from the \fct{summary} object:
\begin{Schunk}
\begin{Sinput}
R> est_lambda <- summary(ZABCNO_educ)$lambda[1]
R> se_lambda  <- summary(ZABCNO_educ)$lambda[2]
R> (ci_lambda <- est_lambda + c(-1, 1) * qnorm(0.975) * se_lambda)
\end{Sinput}
\begin{Soutput}
[1] -0.2588786 -0.2588234
\end{Soutput}
\end{Schunk}
The resulting confidence interval for $\lambda$ is entirely negative and excludes zero, providing statistical evidence to reject $\mathcal{H}_0$. To visually assess the overall goodness-of-fit of both models, we can examine the simulated envelope plots now obtained via the \fct{plot} method in the \CRANpkg{BCSreg} package. We generate the plots for both the ZABCS and the zero-adjusted log-symmetric (\code{ZALS\_educ}) fits as follows.
\begin{Schunk}
\begin{Sinput}
R> # Envelope plot for the general ZABCS model
R> plot(ZABCNO_educ, which = 4)
R> # Envelope plot for the restricted log-symmetric model (lambda = 0)
R> plot(ZALS_educ, which = 4)
\end{Sinput}
\end{Schunk}

\begin{figure}[htbp]
   \centering
   \includegraphics[width=0.48\textwidth]{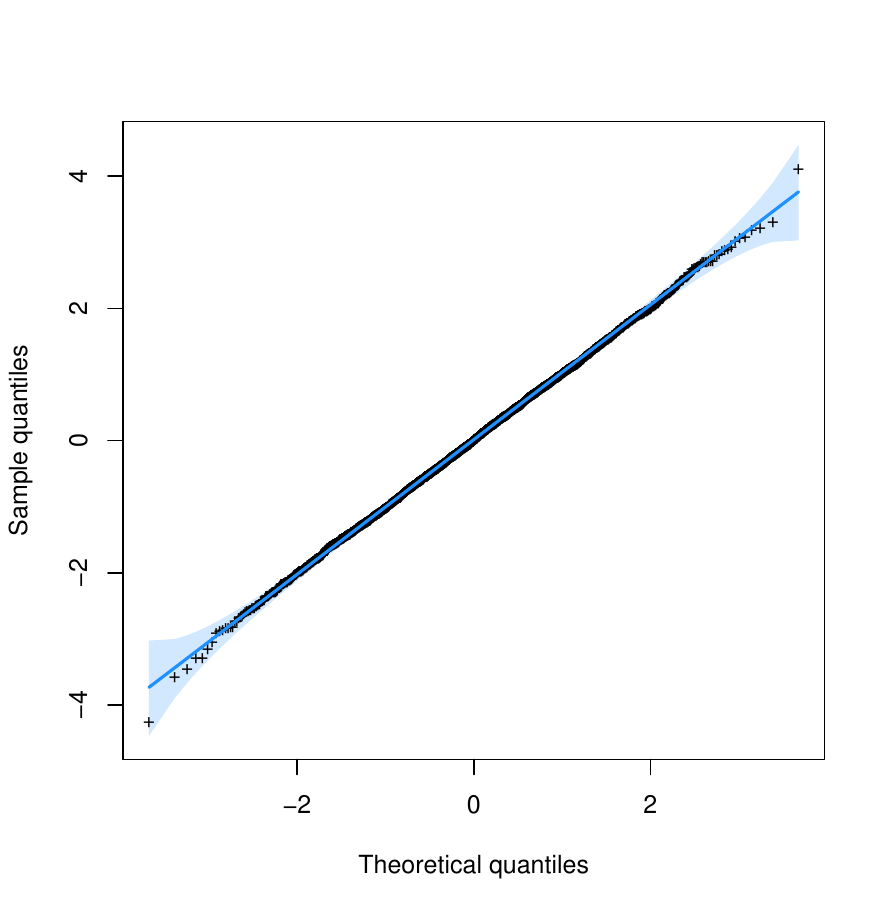}
   \includegraphics[width=0.48\textwidth]{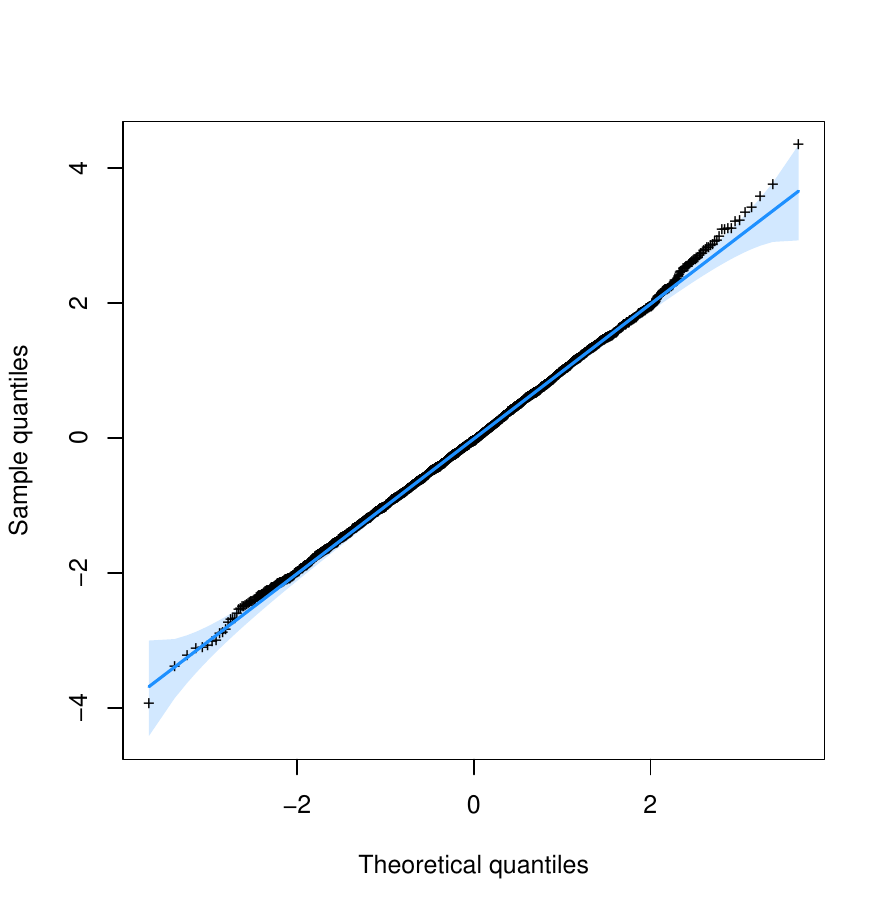}
   \caption{Simulated envelopes for the randomized quantile residuals of the ZABCS model (left) and the zero-adjusted log-symmetric model (right).}
   \label{fig:envelopes_educ}
 \end{figure}

The resulting plots, presented in Figure \ref{fig:envelopes_educ},  indicate that the quantile residuals from both models are well-behaved, with the vast majority of points falling within the confidence bands. At a visual level, both models appear to provide a satisfactory fit.

The \CRANpkg{BCSreg} package allows the user to specify different link functions for the zero-adjustment submodel (i.e., the probability of observing a strictly zero expenditure). To illustrate this capability, we fit the zero-adjusted log-symmetric model using the \code{probit} and \code{cloglog} link functions and compare them against the default \code{logit} link. We evaluate the fits by extracting the Akaike Information Criterion (AIC) and the Bayesian Information Criterion (BIC) as follows:
\begin{Schunk}
\begin{Sinput}
R> measures <- sapply(c("logit", "probit", "cloglog"), function(x) {
+    fit <- update(ZABCNO_educ, link.zero = x)
+    round(c(AIC(fit), BIC(fit)), 1)
+  })
R> rownames(measures) <- c("AIC", "BIC")
R> measures
\end{Sinput}
\begin{Soutput}
     logit probit cloglog
AIC 6698.8 6706.6  6714.5
BIC 6775.0 6782.9  6790.7
\end{Soutput}
\end{Schunk}

In this application, the default \code{logit} link function yields the lowest values for both the AIC and BIC, indicating that it provides the most adequate fit for the probability of zero expenditures. Conversely, the \code{cloglog} link presents the highest information criteria, making it the least favorable choice among the three options. This straightforward comparison highlights how easily users can explore and select alternative link functions within the \CRANpkg{BCSreg} framework to optimize their model specifications.

\section{Concluding remarks} \label{sec:summary}

In this paper, we introduced the \CRANpkg{BCSreg} package, a comprehensive R framework with a collection of tools for regression analysis of non-negative data, including strictly positive and zero-inflated observations, based on the class of the Box-Cox symmetric (BCS) distributions and its zero-adjusted extension. The BCS distributions are a class of flexible probability models capable of describing different levels of skewness and tail-heaviness. The package offers a comprehensive regression modeling framework, including estimation and tools for evaluating goodness-of-fit. Currently, the package includes eight  distributions in the BCS class. The applications in the previous sections illustrate the ability of the package to fit different BCS regression models, including the log-symmetric and zero-adjusted models.

\section*{Acknowledgments}

The first author gratefully acknowledges partial funding from the São Paulo Research Foundation (FAPESP), Brazil (Grant 2024/08343-9).

\bibliography{refs}

\address{
Francisco F. Queiroz\\
  Department of Statistics\\
  Institute of Mathematics, Statistics and Computer Science\\
  University of S\~ao Paulo\\
  Rua do Mat\~ao, 1010\\
  05508-090, S\~ao Paulo, S\~ao Paulo, Brazil\\
  E-mail: \email{felipeq@ime.usp.br}
  }

\address{
Rodrigo M. R. de Medeiros\\
  Department of Statistics\\
  Federal University of Rio Grande do Norte\\
  Av. Senador Salgado Filho, 3000\\
  59078-900, Natal, Rio Grande do Norte, Brazil\\
  E-mail: \email{rodrigo.matheus@ufrn.br}
  }
\end{article}

\end{document}